\documentclass[aps,twocolumn,prl]{revtex4}
\usepackage[hypertex,linkcolor=red]{hyperref}
\usepackage{exscale}
\usepackage{graphicx}
\usepackage{amsmath}
\usepackage{latexsym}
\usepackage{amsfonts}
\usepackage{amssymb}

\DeclareMathOperator{\Tr}{tr}

\begin{document} 

\title{Distributed quantum dense coding}

\author{
D. Bru{\ss}$^1$,  G.M. D'Ariano$^2$, M. Lewenstein$^1$, C. Macchiavello$^2$,
A. Sen(De)$^1$, and U. Sen$^1$ }

\affiliation{$^1$Institut f\"ur Theoretische Physik, 
Universit\"at Hannover, D-30167 Hannover,
Germany\\
$^2$Dipartimento di Fisica ``A. Volta" and INFM-Unit\'a di Pavia,
Via Bassi 6, 27100 Pavia, Italy}

\begin{abstract}

We introduce the notion of distributed quantum dense coding, i.e.
the generalization of quantum dense coding to more than one sender
and more than one receiver. We show that global operations 
(as compared to local operations) of the senders
do not increase the information transfer capacity, 
in the case of a single receiver. 
For the case
of two receivers, using local operations and classical
communication, 
a non-trivial upper bound 
for the capacity is derived. We propose a general classification 
scheme of quantum
states according to their usefulness for dense coding. In the bipartite
case (for any dimensions), bound entanglement is not useful for this task.

\end{abstract}

\maketitle

Entanglement is considered to be 
the most important resource for quantum information
\cite{Bouwmeesterbook},
 as it allows for new quantum protocols
such as superdense coding, quantum teleportation and quantum
cryptography. It is therefore of great importance to classify quantum 
states according to their entanglement properties, in particular
with respect to their usefulness for a given quantum information task.
An important example of such classification concerns the distillability
of quantum states, i.e. the question whether entanglement can be 
concentrated by local operations \cite{reflectionsML}.
Recently, the question of usefulness of states for quantum teleportation
\cite{HHH_fidelity}
and quantum cryptography
 \cite{CurtyMLL}
has been addressed.

In this Letter, we 
introduce the general concept of distributed dense coding
(see also \cite{quant-ph/0402089}) and present
a classification of mixed states according to 
their {\em dense-codeability}
(DC).
The idea of dense coding is  to use previously shared entanglement 
between a sender  and a receiver, to  send more
 information than that is possible without the resource of entanglement.
We establish a full DC-classification for two-party systems,
generalizing Refs. \cite{Plenio, Hiroshima, Buzek-Ziman}. In particular,
 we show that 
bipartite bound entangled states, in any dimensions, cannot be used for dense coding.
Furthermore, we consider 
 the case of {\em several} senders and receivers,
in three different scenarios: (i) the senders/receivers are distant and not
allowed to communicate among themselves,
(ii) they can use local operations and classical communication (LOCC),
(iii) they can perform global operations.
We present the classification structure for these scenarios.
For the case of a single receiver, we obtain the exact DC-capacity.
Surprisingly, 
this capacity cannot be increased by communication between the
senders or their joint operations. Moreover, 
states which are bound entangled 
in the senders to receiver cut,
are not useful in this scenario. 
For the case of more than one receiver,
we obtain upper bounds for the corresponding DC-capacities.

Let us first consider the bipartite scenario.
The amount of classical information that can be sent via a \(d\)-dimensional 
quantum system is at most \(\log_2d\) bits (binary digits). This is due to 
the Holevo bound
\cite{Holevo}.
In quantum dense coding, entanglement
 between the sender and receiver allows 
to go beyond this bound \cite{BW}.
 If the sender and receiver -- hereafter
called Alice (A) and Bob (B) --  share an entangled bipartite state in 
\(d_A \otimes d_B\), 
Alice  is sometimes able to send more than 
\(\log_2 d_A\) bits to Bob, i.e. more than the maximal information
content of her subsystem without  any shared entanglement.
However, she
 certainly cannot
 send more than \(\log_2d_A + \log_2 d_B\) bits to Bob,
 as required by the Holevo bound.

Given a previously shared state \(\rho^{AB}\) 
in dimension \(d_A \otimes d_{B}\),
a general dense coding protocol
consists of two steps.

1. 
 Alice performs a local unitary transformation \(U_i\) 
with probability \(p_i\) on her part of
 \(\rho^{AB}\). This means that
she transforms the state 
\(\rho^{AB}\) to the ensemble \(\{p_i, \rho_i^{AB}\}\), where
\(\rho_i^{AB} = U_i \otimes I_{d_B} \rho^{AB} U_i^{\dagger}\otimes I_{d_B}\).
Here
\(I_{d_B}\) 
is the identity operator 
on Bob's Hilbert space.  
Alice then sends her part of the ensemble 
state 
to Bob.  

2. Bob  extracts  the maximal information about the index 
\(i\) from the ensemble \(\{p_i, \rho_i^{AB}\}\),
where now the total state is at his side,
 by performing 
suitable measurements.

The maximum amount of
information that Bob can gather  from his measurement is bounded 
from above by the Holevo quantity \cite{Holevo}
\begin{equation}
\label{Holevo_bound}
S(\overline{\rho}) - \sum_i p_i S(\rho_i^{AB}) 
= \sum_i p_i S(\rho_i^{AB} \parallel
\overline{\rho}).
\end{equation} 
Here \(S(\varsigma)= - \mbox{tr}(\varsigma \log_2 \varsigma)\) denotes the von Neumann entropy,
\( S(\varrho \parallel \varsigma)\) 
\( =\mbox{tr} (\varrho \log_2 \varrho - \varrho \log_2 \varsigma)\)
is the relative entropy, and 
\(\overline{\rho} = \sum_i p_i \rho_i^{AB}\). 
This bound can be attained asymptotically \cite{Schumacher123},
so that
the capacity of dense coding is defined as
\(
 \chi =  \max \sum_i p_i S(\rho_i^{AB} \parallel \overline{\rho})
\),
where the maximization is over all sets \(\{U_i\}\) of unitaries  performed by Alice,
and all choices of probabilities \(\{p_i\}\).

For \(d_A\otimes d_B\) systems, with \(d_A = d_B=d\), it was shown in Ref. 
\cite{Hiroshima} that the maximum
is reached for a complete set of orthogonal unitary operators
$\{W_j\}$, sampled with equal probabilities, 
and obeying 
the {\em trace rule} 
$\frac{1}{d^2}\sum_j W_j^\dag \Xi W_j=\Tr[\Xi]I$, for any operator
$\Xi$. A typical example of such a set is provided by the group of
 shift-and-multiply operators 
\(
W_{(p, q)} \left|j\right\rangle = \exp\left(\frac{2 \pi p j }{d}\right) 
\left|j + q (\mbox{mod  } d) \right\rangle
\),
where  \(\{\left|j\right\rangle\}\) denotes an orthonormal basis and
 \(p,q,j = 0, \ldots, d-1\).

 In a similar way one can show that the same sets of unitary operators
with equal probabilities are also optimal for bipartite systems with \(d_A \ne d_{B}\). 
Let us give a brief outline of the proof. As in 
Ref. \cite{Hiroshima}, the optimization of 
the dense coding capacity  proceeds in  three steps.

{\em Step 1.} 
The average state of the ensemble \(\{\frac{1}{d_A^2}, \rho_j\}\),
 that is obtained  after 
Alice performs the unitary transformations \(W_j\)  
on her subsystem,
is 
\(
\overline{\rho}{'} = \frac{1}{d_A} I_{d_A} \otimes \rho^B 
\),
where \(I_{d_A}\) is the identity operator 
on Alice's Hilbert space,
and \(\rho^B = \mbox{tr}_A \rho^{AB}\).
Let \(\chi{'}\) be the capacity for this particular choice 
of unitaries, so that \(\chi{'} = \frac{1}{d_A^2} \sum_j S(\rho_j \parallel \overline{\rho}{'})\).

{\em Step 2.} The capacity $\chi{'}$  is equal to the relative entropy \(S(\sigma_{AB} \parallel \overline{\rho}{'})\), 
for \(\sigma_{AB} = U \otimes I_{d_B} \rho^{AB} U^{\dagger} \otimes I_{d_B}\), 
and an arbitrary unitary transformation \(U\) on Alice's part.

{\em Step 3.} Consider now an arbitrary ensemble \({\cal E} = \{p_i, \rho_i =
U_i \otimes I_{d_B} \rho^{AB} U_i^\dagger \otimes I_{d_B}\}\)
produced by unitary operators \(U_i\) applied (with probability \(p_i\)) by Alice.
Let \(\chi_{{\cal E}}\) be the corresponding capacity,
so that \(\chi_{{\cal E}} = \sum_i p_i S(\rho_{i} \parallel \overline{\rho})\). 
Since \(\chi{'} = S(\rho_i \parallel \overline{\rho}{'})\) for all \(i\) 
(see {\em Step 2}),
we have
\(\chi{'}=  \sum_i p_i S(\rho_i \parallel \overline{\rho}{'})\).
By Donald's identity \cite{Donald123},
\(\chi{'}=  \sum_i p_i S(\rho_i \parallel \overline{\rho}) + S(\overline{\rho} \parallel \overline{\rho}{'})
= \chi_{{\cal E}} + S(\overline{\rho} \parallel \overline{\rho}{'})\), which is \(\geq  
\chi_{{\cal E}}\), as relative entropy is a positive quantity.
So this implies that the complete orthogonal set of unitaries 
\(W_j\),
chosen with 
equal probabilities, is an optimal choice for achieving
the capacity for dense coding in \(d_A \otimes d_B\) systems.     
And consequently the capacity of dense coding for a given shared state \(\rho^{AB}\) 
is given by 
\begin{equation}
\label{taj}
\chi = \log_2d_A + S(\rho^B) - S(\rho^{AB}).
\end{equation}
The quantity \(\chi\) could be increased when Alice and Bob
were allowed to  locally operate on the shared state. 
However, 
an increase of \(\chi\) (e.g. via filtering) 
 would require classical communication between 
them. 
As classical information (which is sent from the sender
Alice to the receiver Bob) is the result of the dense coding protocol,
 we cannot allow them to perform
classical communication to effect a change of the shared state. 

A classical protocol (i.e.
a protocol that does not require a shared  quantum state) can 
 be used by Alice to send at most
\(\log_2 d_A\) bits of classical information. 
A shared quantum state 
is thus said to be useful for dense coding
or {\em dense-codeable} (DC), if the corresponding capacity is
 more than \(\log_2 d_A\). 
From Eq. (\ref{taj}), it is clear that such states are precisely 
those for which 
\(
S(\rho^B) > S(\rho^{AB})
\),
i.e. states which 
are more mixed locally than globally. 
For separable states, this inequality 
is never satisfied \cite{HHHdisorder}. 
We show that even bound entangled states \cite{HHH_bound} in \(d_A \otimes d_B\), i.e. states 
that are entangled, and yet they are not distillable, i.e.
it is not possible to 
obtain maximally entangled states from them by LOCC,
cannot be used for dense coding. For \(d\otimes d\) systems, this was pointed out 
in Ref. \cite{horodeckiQIC}.

Let us first state the reduction criterion  \cite{reduction} 
for detecting distillable states: 
If a state \(\rho^{AB}\) is separable or bound entangled, then 
\(\rho^A \otimes I_{d_B} \geq \rho^{AB}\) 
and \(I_{d_A} \otimes \rho^B \geq \rho^{AB}\). There exist
 distillable states that violate this criterion. 
Any state \(\rho^{AB}\) for which 
\(S(\rho^B) > S(\rho^{AB})\) violates the 
reduction criterion \cite{Wolf} (see also \cite{etao-dewa-uchit}), and is
hence distillable. 
%
Thus,
\(S(\rho^B) > S(\rho^{AB})\)
is not satisfied by {\em any}
 bound entangled state:
Bipartite bound entanglement
is not useful for dense coding.
Note also that one  cannot 
use a bound entangled state either
to obtain a higher fidelity
than classically, 
in a teleportation protocol \cite{teleportation, HHH_fidelity}.

This concludes our studies of bipartite dense coding, where the
capacity for any given composite state is described by
Eq. (\ref{taj}). Note that  any {\em pure} entangled bipartite
state is useful for dense coding, whereas there exist mixed
entangled states, even in 
dimension \(2 \otimes 2\), which are not - e.g. a Werner state
with singlet fraction less than \( \approx .7476\).
By contrast, all 
entangled states in \(2 \otimes 2\) and \(2 \otimes 3\) are useful for 
teleportation \cite{HHH_fidelity}. 
This shows that teleportation and dense coding are {\em inequivalent}
tasks.
In higher dimensions, at least the
distillable states that violate the 
reduction criterion  \cite{reduction}, are useful for teleportation.
This is because, states that violate the reduction criterion, 
either already have nonclassical teleportation fidelity, 
or can be transformed into such a state, by single-side
single-copy filtering operations \cite{HHH_fidelity}.  
Moreover, DC states 
violate the 
reduction criterion
\cite{Wolf}, and hence 
are useful for teleportation. 
\begin{figure}[h]
  \includegraphics[width=6cm]{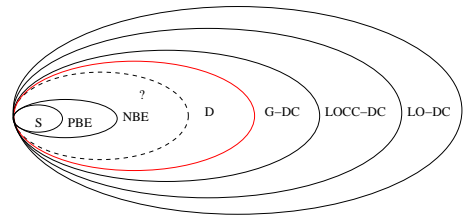}
    \caption{Classification of {\em multipartite} quantum states,
according to their usefulness for dense coding with more than one receiver. 
S, PBE, NBE, D stand respectively for separable, 
bound entangled states with 
positive partial transpose, bound entangled states with nonpositive 
partial transpose (if existing), distillable non-G-DC states (each 
with respect to the bipartite split between the senders and receivers); 
see text for other notations.
For a single receiver in the multiparty case, and for 
bipartite systems, there are shells for S, PBE, NBE, 
D, G-DC only. 
The NBE to D boundary is not convex, provided a certain NBE state 
exists \cite{SST}, while the convexity of 
G-DC to LOCC-DC boundary remains 
an 
open
 problem.  Other boundaries are convex. In particular, the 
 convexity of the D to G-DC boundary follows from \cite{Wehrl}.
}
 \label{figure1}
\end{figure}

We will now consider a scheme of dense coding for multipartite states,
starting with the case of a single receiver.
Suppose that there are \(N-1\) Alices, say, 
\(A_1, A_2, \ldots A_{N-1}\) and a single Bob (B). The Alices want to send (classical) information to Bob. 
The information of one Alice will in general be different from another Alice. 
To do this, they use a previously shared \(N\)-party state 
\(\rho^{A_1 \ldots A_{N-1}B}\). To start,
the \(j\)th Alice \(A_j\)
chooses the unitary tranformation \(U_{i_j}^{A_j}\) with probability \(p_{i_j}^{A_j}\), and applies it on her 
part of the 
state \(\rho\).
After performing the unitary transformations, the Alices send their respective parts to Bob. 
Then Bob  makes a global measurement on the total 
system, to gather 
maximal information
about Alices' ensemble.
Here,
 Bob has no restriction in optimizing over the global 
measurement, and the Holevo quantity
is defined by Alices' action.
Note that the Holevo bound can be achieved asymptotically for 
{\em product} encodings of the
signal states \cite{Schumacher123}. Therefore 
it can be reached asymptotically also in the
present case of many Alices at distant locations. From the complete orthogonal set 
$\{W^{A_l}_{j_l}\}$ for $A_l$ we can construct the set of local operators 
$\otimes_l W^{A_l}_{j_l}$ which is a complete and
 orthogonal set for the composite system of all
Alices, whence the trace rule holds for their global Hilbert space. Then, the situation is
equivalent to the previous case of a single Alice. Using {\em Steps 2, 3}, 
discussed for   
 \(d_A \otimes d_B\) systems,
it follows
that the capacity of 
distributed dense coding with a single 
receiver is 
\begin{eqnarray}
\label{Ashok}
 \chi^{A_1 \ldots A_{N-1}B}&=& 
  \log_2 d_{A_1} + \ldots + \log_2 d_{A_{N-1}} \nonumber \\ 
&& + S(\rho^B)
 - S(\rho^{A_1 \ldots A_{N-1}B}). 
\end{eqnarray}
Notice that
the right hand side  of Eq. (\ref{Ashok}) 
is equal
to the capacity of dense coding when the Alices are together 
(see Eq. (\ref{taj})).
We have thus shown  the  surprising fact that
the Alices do \emph{not} need to perform global unitaries 
 to attain the maximal capacity in a dense coding protocol. We conclude that, 
also in the present scenario, a state 
which is bound entangled in the bipartite cut \(A_1 \ldots A_{N-1} : B\),
cannot be used for dense coding, since analogous
 considerations as before show that one cannot have 
\(S(\rho^B) >
  S(\rho^{A_1 \ldots A_{N-1}B})\) for such a state.

We now consider the situation of several senders (called Alices, 
\(A_1\), \(\ldots\), \(A_{N-1}\))
 and two receivers (called Bobs, \(B_1\), \(B_2\)). 
If the receivers are distant and do not communicate, the corresponding
DC-capacities are simply additive. This case is denoted in Fig. \ref{figure1} as 
LO-DC.
Let us therefore study the case where the 
Bobs are far apart, 
but are allowed to use  
LOCC between them,  denoted as LOCC-DC in Fig. \ref{figure1}. 
Here,
some of the Alices, say \(A_1, \ldots, A_k\), send their parts of the shared 
state \(\rho^{A_1 \ldots A_{N-1} B_1 B_2}\) to \(B_1\), while the rest of
 the Alices, 
\(A_{k+1}, \ldots, A_{N-1}\), send their states to \(B_2\).
Finally, \(B_1\) and \(B_2\) share the ensemble \(\{r_i, \zeta_i\}\), 
given by 
\(
 r_i = p_{i_1}^{A_1} \ldots p_{i_{N-1}}^{A_{N-1}}\), 
\(
 \zeta_i =
U_{i_1}^{A_1} \otimes \ldots \otimes U_{i_{N-1}}^{A_{N-1}} \otimes I_{d_{B_1}} \otimes I_{d_{B_2}}
 \rho^{A_1  \ldots A_{N-1} B_1 B_2} 
U_{i_1}^{A_1\dagger} \otimes \ldots \otimes U_{i_{N-1}}^{A_{N-1}\dagger} \otimes I_{d_{B_1}} \otimes I_{d_{B_2}}
\),
where the unitary operator \(U_{i_j}^{A_j}\) is applied by \(A_j\) with probability \(p_{i_j}^{A_j}\).
Note that \(B_1\) and \(B_2\) are allowed to apply LOCC in the bipartite cut 
\(A_1 \ldots A_k B_1 : A_{k+1} \ldots A_{N-1} B_2\).

Let us denote the classical information 
that can be obtained by the Bobs
in this setting as \(I_{acc}^{LOCC}\).
Its asymptotic version, 
maximized over all
choices of unitaries and probabilities by the Alices, is the DC-capacity 
in this 
case,  denoted as \(\chi^{LOCC}\).
%
%
%
%
%
%
%
%
%
A Holevo-like universal upper bound for \(I_{acc}^{LOCC}\),
valid also for its  asymptotic version, is known \cite{PMAU}. 
In the present case, it  reads
\(
\chi^{LOCC} \leq \max \left( S(\overline{\zeta}^{(1)}) + S(\overline{\zeta}^{(2)})
                                      - \max_{x=1,2} \sum_i p_i S(\zeta_i^{(x)}) \right)
\),
where 
\(\overline{\zeta}^{(1)} = \Tr_{A_{k+1} \ldots A_{N-1} B_2} \overline{\zeta}\),
\(\overline{\zeta}^{(2)} = \Tr_{A_{1} \ldots A_{k+1} B_1} \overline{\zeta}\), with 
\(\overline{\zeta} = \sum_i r_i \zeta_i\), and  \({\zeta}_i^{(1)} = \Tr_{A_{k+1} \ldots A_{N-1} B_2} {\zeta}_i\),
\({\zeta}_i^{(2)} = \Tr_{A_{1} \ldots A_{k+1} B_1} {\zeta}_i\).
The unspecified maximization 
is over all choices of 
unitaries 
and probabilities by the  Alices.

To obtain a more useful bound, note that for any bipartite state
 \(\varrho^{AB}\), 
local unitaries cannot change the spectrum of the global as well as the
 local density matrices.
In particular, for arbitrary unitaries \(U^A\) and \(U^B\) acting 
on \(\varrho^{AB}\) to obtain 
\(\varrho{'}^{AB} = U^A \otimes U^B \varrho^{AB} U^{A\dagger} \otimes
 U^{B\dagger}\), 
we have
 \(S(\Tr_B \varrho^{AB}) = S(\Tr_B \varrho{'}^{AB})\) and 
\(S(\Tr_A \varrho^{AB}) = S(\Tr_A \varrho{'}^{AB})\).
Using this fact, the 
bound on \(\chi^{LOCC}\)
can be 
simplified to obtain
\(
\chi^{LOCC} \leq \max \left( S(\overline{\zeta}^{(1)}) + S(\overline{\zeta}^{(2)}) 
\right)
                                      - \max_{x=1,2} S(\rho^{(x)})
\), 
where \({\rho}^{(1)} = \Tr_{A_{k+1} \ldots A_{N-1} B_2} {\rho}\),
\({\rho}^{(2)} = \Tr_{A_{1} \ldots A_{k+1} B_1} {\rho}\),
 and the unspecified maximization is as before.
This maximization can be performed as follows. First, note that
the maximizations for the two subsets of Alices are independent,
as they concern disjoint subspaces of the Hilbert space.
Thus, we have to find the maximum of the concave function \(S(\overline{\zeta}^{(x)})\)
\((x=1,2)\). Moreover, the  
\(\overline{\zeta}^{(x)}\) form a convex set, for all choices of  unitaries 
\(U_{i_j}^{A_j}\) and probabilities \(p_{i_j}^{A_j}\).
Thus to achieve this maximum, it is sufficient to show that the
first derivative of $S$ vanishes, because here
a local maximum is the global one. Perturbation of the
 solution from the previous maximization tasks,
namely 
\(W_j\)
with equal probabilities, shows 
in a straightforward way that
this solution is again the optimal one.
Thus, we arrive at 
\begin{eqnarray}
&& \chi^{LOCC} \leq
\log_2 d_{A_1}   + \ldots + \log_2 d_{A_{N-1}} \nonumber \\
&&+ S(\rho^{B_1}) + S(\rho^{B_2}) 
- \max_{x=1,2} S(\rho^x) \equiv 
{\cal B}^{LOCC}, 
\label{blocc} 
\end{eqnarray}
where 
\(\rho^{B_1} = \Tr_{A_1 \ldots A_{N-1} B_2} \rho  \) and \(\rho^{B_2} = 
\Tr_{A_1 \ldots A_{N-1} B_1} \rho  \). 
Analogous arguments as in \emph{Steps 2}, \emph{3} also give Eq. (\ref{blocc}).
%

A trivial lower bound on \(\chi^{LOCC}\) is given by the case where
the two Bobs do not use communication; thus their two channels are
independent, and the capacities add. We denote the capacity
without communication as $\chi^{B_1B_2}$, and thus have
\(
\chi^{LOCC} \geq \chi^{B_1} + \chi^{B_2} = \chi^{B_1B_2}
\).
A trivial upper bound on \(\chi^{LOCC}\) is obtained by using the fact that
 the Bobs can 
 obtain more (at least, not less) information, if they are 
together and are allowed to use global measurements, referred to as G-DC in Fig. 
\ref{figure1}. 
Let us call this bound the global DC-capacity
 $\chi^{glob}$:
\(
\chi^{LOCC} \leq \log_2 d_{A_1}   + \ldots + \log_2 d_{A_{N-1}}  +  
S(\rho^{B_1B_2}) -  S(\rho)
 = 
\chi^{glob}
\).
We summarize our results for the  dense-codeability 
of a given multipartite quantum state, for two receivers,  in 
Fig. \ref{figure1}.
We call a state dense-codeable, if its capacity is greater than
$\log_2 d_{A_1}   + \ldots + \log_2 d_{A_{N-1}}$, and 
locally dense-codeable if \(\chi^{B_1 B_2}> \log_2 d_{A_1}   + \ldots + \log_2 d_{A_{N-1}}\). 


We now provide examples for  the sets indicated in Fig.
 \ref{figure1}, and thus show that the sets  are non-empty. 
Note also, that for these examples of DC
states,
 one can add the identity
to the corresponding state (up to a certain limit), and still keep the noisy
state dense-codeable. Therefore the sets are not of measure zero.

An example of a state that is G-DC, but not LOCC-DC
(i.e. 
${\cal B}^{LOCC}\leq\log_2 d_{A_1}   + \ldots + \log_2 d_{A_{N-1}} < \chi^{glob}$)
is 
\(
\frac{1}{2} \left( \left|0000\right\rangle + \left|0101\right\rangle +
\left|1000\right\rangle + \left|1110\right\rangle \right)
\) from \cite{Frank123},
where the first two parties are senders and the last two parties are 
receivers, with the first (respectively, second)
party sending her subsystem to the third (respectively, fourth).

The four-qubit GHZ state \cite{GHZ}, namely 
\((\left|0000\right\rangle + \left|1111\right\rangle)/\sqrt{2}\),
is not locally DC, as the two-party
reduced density matrices are separable. 
However, it is useful for
LOCC-DC:
When the two senders choose the Pauli unitaries with equal probabilities,
one can show that the two receivers can completely distinguish 
the resulting ensemble
of eight orthogonal states by LOCC. This protocol and 
the upper bound in Eq. (\ref{blocc}), give
\(\chi^{LOCC} = 3\). 

A trivial example for a state that is already locally
DC is the tensor product of two singlets. A four-party
$W$ state \cite{W} is not locally DC, but it is yet unknown whether
it is LOCC-DC. The general problem in proving that a state
is useful for LOCC dense coding is that the bound in (\ref{blocc})
is sometimes not very tight, and can even be higher than 
the global bound $\chi^{glob}$, as is the case for the bound
entangled states of Ref. \cite{DuerCiracTarrachstates}. The
question whether there exist multipartite bound entangled states
that are DC remains open.
We point out here that the 
ordering of states that is induced by the task ``dense coding'', as illustrated in  Fig.
 \ref{figure1}, is different from the ordering induced by other 
entanglement criteria, e.g. as described in \cite{Frank123}.
Each quantum information processing objective may even lead to
its own structure of quantum states.

Finally, 
it is formally possible to generalise these considerations 
to the case where there are more than two receivers.
However, the main obstacle is that there is as yet no 
good estimation of mutual information that is accessible locally, 
for the case of more than two parties. 
For an attempt in this direction, see
Ref. \cite{quantumcorrelation}. 

In summary,
we have introduced the notion of dense-codeability, i.e. the
usefulness of a given quantum state for dense coding. We have
generalized bipartite dense coding to the multipartite case, 
and investigated the classification of  entangled states  according
to their dense codeability. We
have presented a full classification for the bipartite case, and showed
that here
bound entangled states in any dimensions are not dense-codeable. In the multipartite case the
capacity of dense coding depends  on the 
possibility of interactions between
the receivers. Here, we proposed a classification scheme and
showed examples for the various identified classes.

We acknowledge support of the Deutsche Forschungsgemeinschaft 
(SFB 407, SPP 1078), the 
Alexander von Humboldt  
Foundation, and
the EC Contract No. IST-2002-38877 QUPRODIS.
D.B and M.L. are grateful  for the
hospitality of the organisers of the Pavia Workshop
{\em Quantum Information Processing and Quantum Communications,
} supported by the 
ESF and Ministero dell'Istruzione, dell'Universit\`a e della Ricerca (Cofinanziamento 2002).


\begin{thebibliography}{99}


\bibitem{Bouwmeesterbook}D. Bouwmeester, A. Ekert, and A. Zeilinger (Eds.), 
\emph{The Physics of quantum information} (Springer, Berlin, 2000);
D. Heiss (Ed.), \emph{Fundamentals of Quantum Information,
Quantum Computation, Communication, Decoherence and All That}
(Springer, NY, 2002).



\bibitem{reflectionsML} D. Bru\ss\ {\it et al.}, 
J. Mod. Opt. \textbf{49}, 1399 (2002). 
 


\bibitem{HHH_fidelity} N. Linden and S. Popescu,
Phys. Rev. A  \textbf{59}, 137 (1999);  M. Horodecki, P. Horodecki, and R. Horodecki, 
Phys. Rev. A  \textbf{60}, 1888 (1999). 

\bibitem{CurtyMLL} D. Bru\ss\ {\it et al.}, 
Phys. Rev. Lett. \textbf{91}, 097901 (2003);
M. Curty, M. Lewenstein, and N. L{\"u}tkenhaus, 
\emph{ibid.} \textbf{92}, 217903 (2004);
A. Acin and N. Gisin, quant-ph/0310054.


\bibitem{quant-ph/0402089} X.S. Liu, G.L. Long, D.M. Tong, and F. Li, Phys. Rev. A \textbf{65}, 022304 (2002).


\bibitem{Plenio} S. Bose, M.B. Plenio, and V. Vedral, quant-ph/9810025.

\bibitem{Hiroshima} T. Hiroshima, quant-ph/0009048.


\bibitem{Buzek-Ziman} M. Ziman and V. Bu{\v z}ek, Phys. Rev. A \textbf{67},
042321 (2003). 




\bibitem {Holevo}
J.P. Gordon, in \emph{Proc. Int. School Phys. ``Enrico Fermi,
Course XXXI''}, ed. P.A. Miles, pp. 156 (Academic Press, NY 1964); 
L.B. Levitin, in \emph{Proc. VI National Conf. Inf. Theory, Tashkent}, pp. 111
(1969); A.S. Holevo, Probl. Pereda. Inf. \textbf{9}, 3 1973 [Probl. Inf.
Transm. \textbf{9}, 110 (1973)].

\bibitem{BW} C.H. Bennett and S.J. Wiesner, Phys. Rev. Lett. \textbf{69}, 2881 (1992).





\bibitem{Schumacher123} B. Schumacher and M.D. Westmoreland, Phys. Rev. A \textbf{56}, 131 (1997); 
A.S. Holevo, IEEE Trans. Inf. Theory \textbf{44}, 269 (1998).






\bibitem{Donald123} M.J. Donald, Math. Proc. Cam. Phil. Soc. \textbf{101}, 363 (1987).

\bibitem{HHHdisorder} R. Horodecki, P. Horodecki, and M. Horodecki,
Phys. Lett. A  \textbf{210}, 377 (1996).


\bibitem{HHH_bound} M. Horodecki, P. Horodecki, and  R. Horodecki, Phys. Rev. Lett.
\textbf{80}, 5239 (1998).





\bibitem{horodeckiQIC} M. Horodecki \emph{et al}.,
QIC, \textbf{1}, 70 (2001).

\bibitem{reduction} M. Horodecki and P. Horodecki, Phys. Rev. A \textbf{59}, 4206 (1999);
N.J. Cerf, C. Adami, and R.M. Gingrich, \emph{ibid}.
\textbf{60}, 898 (1999).


\bibitem{Wolf} K.G.H. Vollbrecht and M.M. Wolf, quant-ph/0202058.



\bibitem{etao-dewa-uchit} T. Hiroshima, Phys. Rev. Lett. \textbf{91}, 057902 (2003).
\bibitem{teleportation} C.H. Bennett {\it et al.}, 
Phys. Rev. Lett. \textbf{70}, 1895 (1993).



 




\bibitem{SST} P.W.  Shor, J.A. Smolin, and B.M. Terhal,
Phys. Rev. Lett. \textbf{86}, 2681 (2001).



\bibitem{Wehrl} A. Wehrl, Rev. Mod. Phys. \textbf{50}, 221 (1978).



\bibitem{PMAU} P. Badzi{\c a}g, M. Horodecki, A. Sen(De),  and U. Sen, Phys. Rev. Lett. 
\textbf{91}, 117901 (2003).




\bibitem{Frank123} F. Verstraete, J. Dehaene, B. De Moor, and H. Verschelde,  
Phys. Rev. A \textbf{65}, 052112 (2002).


\bibitem{GHZ} D.M. Greenberger, M.A. Horne, and A. Zeilinger, in \emph{Bell's
 Theorem, Quantum Theory, and Conceptions of the Universe}, ed. M. Kafatos,
 (Kluwer, Dordrecht, 1989).



\bibitem{W}  W. D{\" u}r, G. Vidal, and J.I. Cirac, Phys. Rev. A \textbf{62}, 062314 (2000).
 



\bibitem{DuerCiracTarrachstates}W. D{\" u}r, J.I. Cirac, and R. Tarrach, 
Phys. Rev. Lett. \textbf{83},  3562  (1999).

\bibitem{quantumcorrelation} M. Horodecki, A. Sen(De), and U. Sen, quant-ph/0310100. 



\end{thebibliography}
\end{document}